\documentclass[preprint2]{aastex}
\usepackage{color}
\usepackage {lineno}
\usepackage [breaklinks=true]{hyperref}
\usepackage{breakcites}

\shorttitle{[Sr/Ba] and [Sr/Eu] Dependence on the EOS}
\shortauthors{Famiano et al.}

\begin{document}
%\linenumbers
\date{\today}
\title{Dependence of the Sr-to-Ba and Sr-to-Eu Ratio on the Nuclear Equation of State
in Metal Poor Halo Stars}
\author{M.A. Famiano\altaffilmark{1}}
\affil{Dept. of Physics and Joint Institute for Nuclear Astrophysics, Western Michigan University, 
1903 W. Michigan Avenue, Kalamazoo, MI 49008-5252, USA}
\email{michael.famiano@wmich.edu}

\author{T. Kajino\altaffilmark{2}}
\affil{National Astronomical 
Observatory of Japan, 
2-21-1 Mitaka, Tokyo 181-8588 Japan}
\email{kajino@nao.ac.jp}
%\author{R.N. Boyd}
%\affil{National Astronomical 
%Observatory of Japan, 
%2-21-1 Mitaka, Tokyo, 181-8588, Japan}
%\email{richard11boyde@comcast.net}
\author{W. Aoki}
\affil{National Astronomical 
Observatory of Japan, 
2-21-1 Mitaka, Tokyo 181-8588 Japan}
\and
\author{T. Suda}
\affil{The Research Center for the Early Universe (RESCEU);
	Univ. of Tokyo, 7-3-1 Hongo, Bunkyo-Ku, Tokyo 113-0033 Japan   }
\altaffiltext{1}{National Astronomical 
	Observatory of Japan, 
	2-21-1 Mitaka, Tokyo 181-8588 Japan}
\altaffiltext{2}{Dept. of Astronomy, 
Graduate School of Science; Univ. of Tokyo, 7-3-1 
Hongo, Bunkyo-ku, Tokyo 113-0033 Japan}

\begin{abstract}
A model is proposed in which the light r-process element enrichment in
metal-poor stars is explained via enrichment from a 
truncated r-process, or ``tr-process.''  The truncation of the r-process
from a generic core-collapse event followed by a collapse into an accretion-induced
black hole is examined in the framework of a galactic chemical evolution model.
The constraints on this model imposed by observations of extremely metal-poor stars
are explained, and the upper limits in the [Sr/Ba] distributions are found to be
related to the nuclear equation of state in a collapse scenario.  The scatter
in [Sr/Ba] and [Sr/Eu] as a function of metallicity has been found to be consistent with
turbulent ejection in core collapse supernovae.  Adaptations
of this model are evaluated to account for the scatter
in isotopic observables.  This is done by assuming
mixing in ejecta in a supernova event.
\end{abstract}

\maketitle
\section{Introduction}
Historically the general properties of the r-process have been well 
identified \citep{b2fh,woo94,wallerstein97,farouqi09}.
The site of the r-process has been identified as originating either in core collapse
supernovae \citep{woo94,takahashi94,farouqi09}, 
or in mergers of binary neutron stars \citep{neutron}. 
Recent
efforts have called the former site into question, although the 
simulations of the
conditions that occur in the centers of core collapse 
supernovae are yet to be simulated in
three dimensions and incorporating general relativity. Effects that are generally not included 
in such calculations, e.g.,
those that would be imposed by sterile neutrinos \citep{wu14, warren14}, 
could also impact those conclusions.
Some of the recent studies have suggested neutron star mergers as 
a possible site of the r-process,
though they may only result in late galactic production of r-process 
elements \citep{wanajo14, goreily11, neutron}. 
Neutron star mergers, on the other hand, may
be complementary to SNe as an r-process source \citep{qian00, shibagaki15}.  
In particular, the role of multiple r-process sites has been explored including sites
incorporating fission cycling as a contribution to galactic abundances \citep{shibagaki15}.  

In the present paper we study high-resolution spectroscopic data on the logarithmic ratios 
[Sr/Ba] and [Sr/Eu] where [A/B]$\equiv \log(A/B)_* - \log(A/B)_\odot$ for extremely
metal poor stars (EMPs, defined as stars for which [Fe/H] $<$ -3.0). 
These data are known
to exhibit large fluctuations, over two orders of magnitude, for 
different stars at the same metallicity, as their
metallicity, defined by [Fe/H], approaches -3.5.
 
\begin{figure}
	\includegraphics[width=\linewidth]{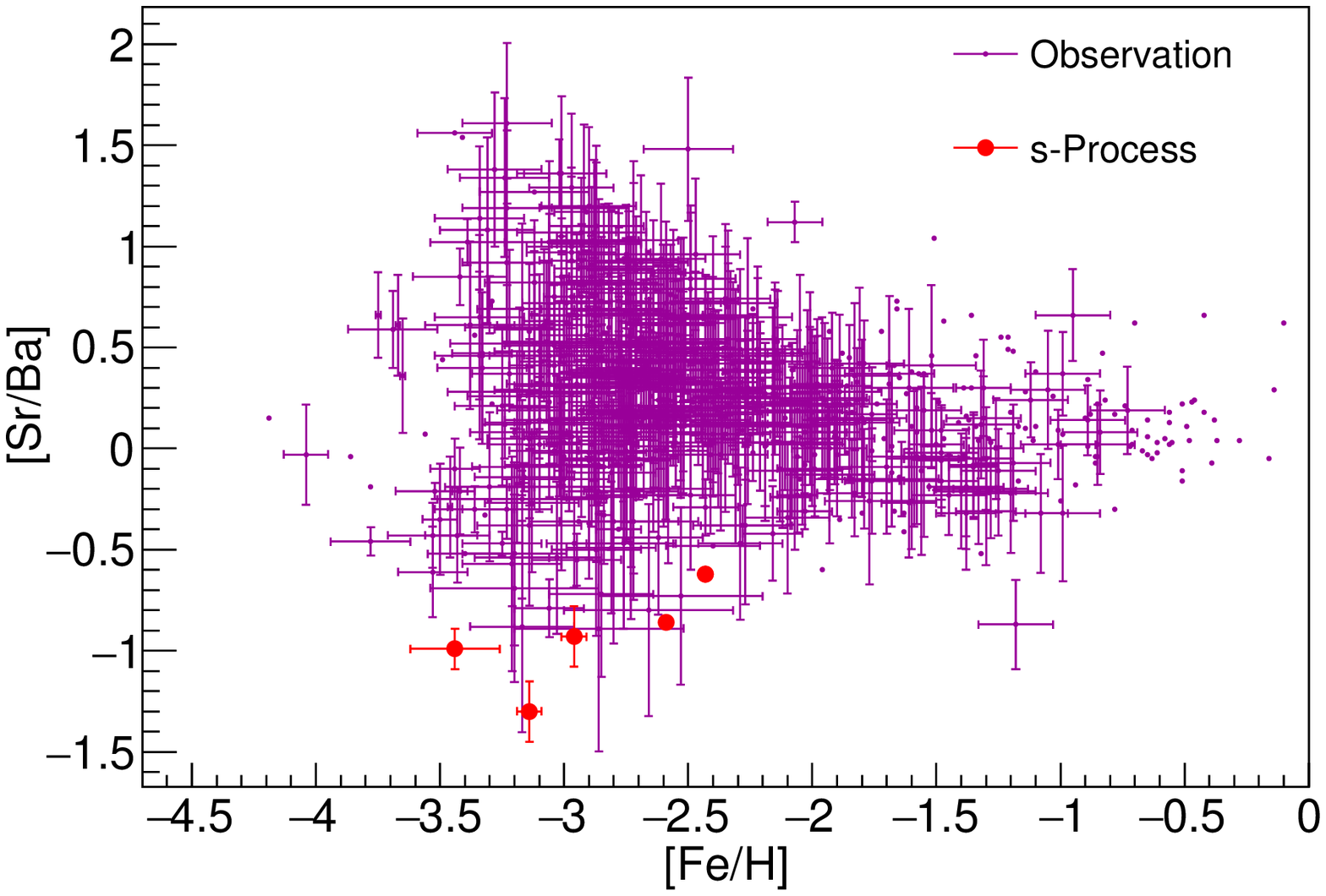}
	\caption{Observational data of metal-poor stars and EMPs showing the scatter
		in [Sr/Ba] as a function of metallicity \citep{saga}.
		Stars 
		that are potentially contaminated by the 
		${\it s}$-process, as discussed in \cite{aoki_apj}; CS 30322-023, CS 29493-090, HE 
		0305-4520, CS 22946-011, CS 22941-005, and CS 
		22950-173 are indicated by filled circles. 
		\label{fig1}}
	%\label{sr_ba_obs}}
\end{figure}

The current status of these data is
indicated in Figure \ref{fig1}. 
The data in that figure were selected from the world's 
collection of
data for EMP stars \citep{saga, suda11, yamada13} for which the abundances of
Ba and Sr were determined from the
same study; these were thought to be the most
 reliable data for these two elements. 
 The selection criterion for these data is the
  same as in \cite{aoki_apj}, but all the
  data are renormalized by the same solar 
  abundances by \cite{asplund09}, based on 
  the recent update of the SAGA database 
  \citep{suda_prep}. In the case of the data selection here, the 
  statistics for these data is described in 
  \cite{aoki_apj}.  The definition of 
  carbon-enhanced stars, [C/Fe] $\geq 0.7$, is 
  the same as in \citet{suda11}(see also \citet{aoki07b}),
  all of which are excluded from the sample. However, the fiducial values of [C/Fe] 
  for individual stars are redetermined by the 
  average of all the reported abundances of 
  carbon abundances after the renormalization by 
  the solar abundances of \citet{asplund09}.
  
		It is particularly interesting to note the sharp decline in
		observations for [Fe/H]$\lesssim$-3.6 \citep{aoki_apj}. In 
		prior work \citep{aoki_apj} this cutoff was noted, and the
		reduced number of observations of stars with observable
		Sr and Ba abundances was pointed out.  Since then, there have
		been more observations of stars with [Fe/H]$<$-3.6 with 
		[Sr/Ba]$>$0.05, though the number still remains low.

We
study these data in the context of two models, 
the Qian-Wasserburg (QW) model \citep{qw} and the
truncated r-process (tr-process) model \citep{bfak}. 
The former QW model is a phenomenological model that
describes the EMP star data in terms of three components. 
This model evaluates the galactic chemical evolution (GCE) of
r-process elements from 
low-mass SNe, normal SNe, and hypernovae (HNe) as the sources
of r-process enrichment in the ISM.  In particular, the
QW model explores the production of Sr and Ba, and 
results from this model are used to examine [Sr/Ba] as a 
function of [Fe/H] in metal-poor stars.  Because the three
sources explored in this model evolve at different rates,
this model makes some progress in explaining the evolution
of [Sr/Ba] in the galaxy.
This
model assumes incomplete mixing of the outputs from the 
progenitor stars which
produced the abundances of the EMP stars being observed; 
the fluctuation in this
mixing is assumed to produce the large spread in the [Sr/Ba] ratio. 
In the tr-process model, the
apparent termination in the [Sr/Ba] data at about [Fe/H] = -3.5 is 
the result of progenitor
stars that exceed the minimum mass for 
direct collapse to black holes (BHs), thereby expelling no r-process 
nuclides to the ISM. Stars with lower masses would have some of 
their r-process production expelled into the ISM. 

The model explored in this paper only attempts to explain the maximum [Sr/Ba] ejected
in a single r-process event; we note that reductions in this ratio may result from mixing
between outer and inner ejecta in the explosion or from asymmetric explosion mechanisms.
We also note that the observed large values of
[Sr/Fe], [Ba/Fe], and [Eu/Fe] (so-called r-II stars)
can be reproduced by turbulent ejection (as
suggested in \cite{aoki_apj}).  The model
here represents one potential avenue for 
producing these extremes.

We explore a previously-proposed model, referred to as the tr-process 
\citep{bfak}, by
which the abundance of
r-process nuclides in very metal-poor stars is the result of
ejection and subsequent enrichment of the interstellar medium (ISM) from a single-event.  The
model assumes an r-process in
a core-collapse scenario which is halted due to an accretion-induced collapse into
a BH or a stalled shock.  In this model, newly synthesized nuclei are only partially emitted into the ISM,
resulting in an abundance distribution enriched in the light r-process elements.  While this 
process is proposed to occur in the neutrino-driven winds of
core collapse supernovae followed by accretion-induced 
collapse, the current model is general enough that it could
occur in other similar events. A possible other site in which
a partial r-process may occur in a collapse scenario is that
of magneto-hydrodynamic (MHD) jets \citep{hidaka15, nakamura15}.  It is thought that these phenomena may be associated with collapsed
massive stars or possibly even hypernovae \citep{nakamura15}.  In one scenario, the ram pressure of the jet is sufficient to allow material to escape the collapse of a massive star (even to a BH) and drive the 
nucleosynthesis within the jet.   The model here is similar in that 
it assumes material escapes the central core in a collapse.     
Furthermore,
the proposed model is not only capable of producing the observed scatter in
[Sr/Ba] and [Sr/Eu] at very low metallicity, but it is shown to relate the
nuclear equation of state (EOS) to the observed upper limits in [Sr/Ba] and [Sr/Eu].

The scenario described here is used to
	 explain the highest values of [Sr/Ba] and [Sr/Eu] in 
	 halo stars and to ascribe a physical explanation to 
	 these values and their possible extremes.  
	 Here, massive stars which produce type II supernovae (SNII) may collapse due to subsequent fallback.  Regardless of the 
	ultimate fate of the collapse, the ensuing fallback can halt the ejection of mass layers deeper in the star.  These layers are the ones that produce the heavier
	r-process elements.  Only the outer layers are ejected prior to the collapse; and it is these layers that result in an r-process enriched in the lighter r-process elements. In this model, the layers that are successfully ejected are determined by the collapse time; only layers which have been ejected prior to the collapse result in ejection of r-process elements into the ISM.  The collapse time is, in turn, dependent on the explosion energy, which is directly dependent on the nuclear EOS.   Layers ejected early are enriched in Sr relative to Ba.  Thus, it is proposed that
	the value of [Sr/Ba] is an observable of the nuclear EOS.  Since the more massive stars have shorter
	lifetimes, and since the collapse time decreases with
	progenitor mass, more Sr is expected to be produced earlier in galactic history relative to Ba.

The model proposed here is only phenomenological.  While we
seek to provide an explanation for the extreme [Sr/Ba] values
in metal-poor halo stars, the more quantitative evaluation
will be developed in a future paper.
Though the proposed model is simple, it allows us to address
	the observed values of [Sr/Ba] at very low metallicity in a GCE model.  This model is
	also supported by prior work to address the role of failed superonovea in the galaxy, particularly as
	they relate to the so-called "red supergiant problem" \citep{hidaka15}.  While prior work has
	sought to address the problem of light-element enhancement in metal-poor stars via various processes \citep{arcones11, cristallo15, shibagaki15, travaglio04}, the model here addresses the contribution of
	failed SNe to galactic chemical evolution via the so-called "tr-process" \citep{bfak}.

In \S\ref{collapse_model}, we give an overview of the collapse scenario in which
a tr-process proceeds. 
This is followed by a description of the collapse
and network calculations used to predict abundance distributions in \S\ref{calculations}.  
In this section, we show that the collapse time is dependent on nuclear EOS.
We describe our results in \S\ref{results} with concluding remarks in \S\ref{conclusions}.
\section{Black Hole Collapse Model of the tr-Process}
\label{collapse_model}
In this paper, we examine the possibility of the "tr-process" described in a prior work 
\citep{bfak, famiano13} in light of its possible effects on GCE.  In this model, the r-process is halted
as mass in a type II supernova (SNII) does not escape
fallback onto the suface of the nascent neutron star.  Mass that is ejected undergoes r-processing, while mass that does
not escape the resulting fallback does not contribute to the
total r-process enrichment of the ISM. While the tr-process
discussed in prior work assumes that a BH collapse is what prevents mass shells from being ejected after the collapse, the model may
be applicable to any fallback scenario.

The model described her follows prior work in which the Sr, Ba, and Eu
production in an r-process is directly related to the post-bounce time of
BH formation, $t_{BH}$.  As with the initial tr-process model, 
we assume that the r-process is universal in post-bounce ejection time
of r-processing zones in a type II supernova independent of the mass
of the progenitor star.  As the goal of this work is to examine the sensitivity of the
EOS dependence of the ejection prior to the post-bounce collapse, we
maintain this assumption of universality in this simplified model.  
We use a simple phenomenological model.  A 
"canonical" SN r-process site is assumed in which the mass fraction of an r-process
element as a function of time is the same for all progenitors, regardless of mass.
While this certainly lacks computational rigor, the purpose is to obtain an approximate
evaluation of how cutoffs in a partial or failed explosion may affect galactic
r-process abundances and to gauge the effects of the the nuclear EOS on galactic 
abundances.

\begin{figure}
	\includegraphics[width=\linewidth]{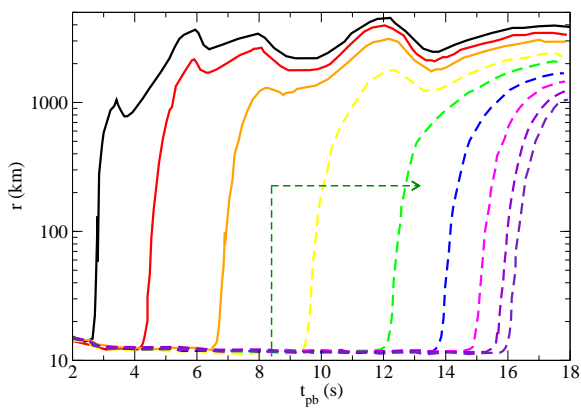}
	\caption{A schematic representation of the tr-process.  
		In the above schematic figure, the collapse time $t_{bh}$ = 8.4 s is indicated by the green vertical line in the figure.  Every shell after the collapse (indicated by the green horizontal arrow) is not 
		ejected.  Successfully ejected shells are those indicated by the solid lines, while those not ejected are indicated by
		the dashed lines.  As an example, trajectories from \cite{woo94} are used, though any explosion model can be used.
		\label{fig2}}
	%		\label{fig2}}
\end{figure}

The model is described
schematically in Figure \ref{fig2}. (In all figures, $t_{pb}$ is any time post-bounce, $t_{bh}$ is the post-bounce time at 
which a collapse - potentially to a black hole - occurs.)
In this model, the radii of mass shells ejected from
a SNII are plotted as a function of time.  Some mass
shells escape the surface of the nascent neutron star and enrich the ISM in r-process elements, while others
fall back to the surface of the star.  In the tr-process, mass shells not ejected are assumed to be those that are not ejected at the time of BH collapse.  
The tr-process as described can have the following effects on the GCE
of Sr and Ba.  More massive non-rotating stars will exhibit fallback and collapse at
earlier times than less massive stars, cutting off ejection of material deeper in
the star in a SN explosion.  This will result in a higher proportion of lighter 
r-process elements ejected into the ISM by more massive stars.  Further, since more
massive stars have shorter lifetimes, higher proportions of lighter r-process elements
are expected to be observed at earlier times in the galaxy, resulting in  
higher [Sr/Ba] values for metal-poor halo stars with lower [Fe/H].
As the galaxy evolves, the values of [Sr/Ba] tend towards the average value observed
on Population I stars.  However, it is the high [Sr/Ba] values that are of interest in this 
model. In addition, the epoch at which those values appeared (corresponding to specific [Fe/H]) is also of interest.  In the tr-process, the collapse time of a star directly depends on 
the nuclear EOS with a stiffer EOS resulting in a more energetic bounce and later collapse time.
A stiffer EOS ultimately results in more Ba ejected into the ISM prior to collapse.   Thus, 
the EOS dictates the time in galactic history that a particular [Sr/Ba] value is produced.  
This could mean that the maximum [Sr/Ba] values are observables of the nuclear EOS.

In prior calculations \citep{bfak,aoki_apj} the Sr, Ba, and Eu
yields in a truncated r-process are directly related to the post-bounce time in an
accretion-induced BH \citep{woo94} or fallback SNII.  The Sr, Ba, and Eu yields in a GCE model were
based on a primary r-process source using existing GCE abundances
where available \citep{ishimaru, travaglio}.  
For the accretion-induced BH the collapse time
was computed based on progenitor mass and metallicity and the nuclear
equation of state \citep{GRID_ref}.  This 
collapse time was then assumed to be the truncation time in a dynamic tr-process. 
Longer truncation times would permit mass shells closer to the proto-neutron star core - which are blown off
later in the
explosion model - to be ejected, either by achieving a velocity in excess of the
escape velocity or perhaps by being ejected in a jet. 
Since these lower-lying ejecta tend to produce heavier r-process nuclides, they enhance the production yields of heavier r-process elements.  The mass-weighted 
yields of Sr and Ba
produced in a tr-process can then be scaled based on the collapse time and ultimately 
on the progenitor mass and metallicity.  In this simplified model, although the collapse time
is a function of the initial mass and metallicity of the progenitor, the scaling of Sr, Ba, and Eu
is a function only of the collapse time \citep{bfak}.

In the r-process, most of the Sr is produced in the early-stage ejecta, while most of the
Eu and Ba is produced in the later-stage ejecta of a type-II supernova (SNII).  Thus, in an explosion model the 
Sr yields in a tr-process
are fairly constant with collapse time and fairly close to their complete r-process values
except for very short collapse times for which essentially no r-process material is ejected.  The Ba
yields, on the other hand, are very small for early and intermediate collapse times and do not
reach their full r-process values until fairly late collapse times (several seconds post-bounce).

In order to relate the collapse time to the progenitor, results from the 
spherical collapse code GR1D\citep{GRID_ref} 
were used to study the relationship between the progenitor star's mass
and metallicity and the Sr, Ba, and Eu yields, which involves the black hole collapse time, for
several nuclear equations of state.  
It is noted that these yields are minimum yields, since
in the non-rotating spherical models employed here, the collapse times
are minimum collapse times.  Longer collapse times may 
result from effects such as asymmetric explosions, rotations, and neutrino heating.

The Sr, Ba, and Eu yields as a function of progenitor mass and metallicity were
calculated in a multi-zone 
GCE model to determine the metallicity
evolution, for example, of [Sr/Ba] \citep{timmes}.  
Results for halo
stars have been determined.  The GCE model parameters employed in 
this study are the same as those used in \cite{timmes}.  
In this model, Sr and Ba yields in massive
stars are based on  the GCE results of \cite{ishimaru} and \cite{travaglio}. 
For a star enriched by the r-process in this model, the Sr and Ba yields are scaled from 
the initial yields of \cite{ishimaru} and \cite{travaglio} according to the mass
and metallicity of the progenitor for each EOS.

We note that there are significant similarities between our 
model and the phenomenological three component QW model \citep{qw}.
Their model has two of their three components competing in the 
EMP star region, with hypernovae dominating for [Fe/H] $<$ -3.5. 
They attribute production only of lighter nuclei, especially Fe, to 
hypernovae, although other studies (e.g. \cite{caballero12,izutani09}) 
suggest that hypernovae might also produce 
a relative enrichment in the lighter r-process elements. The QW model explains the large dispersion of [Sr/Ba] 
data by assuming spatial inhomogeneities in the products of the two 
competing components, and assuming a parameter to define the relative 
contributions of the two components. 

Mathematically the QW model is similar to our model, although ours attempts to
constrain one specific r-process site. The mixing parameters of QW could be used to characterize 
the relative deviations of Sr (light r-process nuclides) and Ba (heavy r-process nuclides) galactic abundances
predicted by this model. 
The two models differ primarily in the sharpness of the 
cutoff of [Sr/Ba] at [Fe/H] $\sim$ -3.5; it would be expected to be more gradual in 
the QW model than in our model.  
%\textcolor{red}{We note that the ratio of the number 
%of stars having [Sr/Ba]$>$0 to those with [Sr/Ba]$<$0 in the metallicity 
%region -3.5$\lesssim$[Fe/H]$\lesssim$-3.0 is 3. If that same ratio applied to the 
%metallicity region -4.0$<$[Fe/H]$<$-3.5, as one might expect would be the case 
%if the mechanisms that produce the stars in the two metallicity regions are 
%the same or are slowly evolving, given the 4 or 5 stars with [Sr/Ba]$<$0 in 
%that region, the number expected for [Sr/Ba] would be 12 to 15. Because that number is
%zero, that might suggest some dramatic change in the production mechanisms 
%for stars having [Sr/Ba]$>$0, [Fe/H]$<$-3.5.}

Another critical feature of the discussion of the predictions of the two 
models lies in the upper
and lower limits on the [Sr/Ba] values that the two models 
impose. In the QW model,
these limits are fixed by the parameter that describes the fraction of the two components
that are assumed to dominate the EMP data. 
These are relatively well
determined functions of [Fe/H]. 

However, the upper and lower limits in the range of [Fe/H] values in the tr-process
model are determined quite differently. The upper limit is essentially infinite,
that is, we will show that stars
might be expected to have huge [Sr/Ba] values, but these would not be included in the
data sample because their Ba abundances would be so low that they would not be detectable.
Thus the upper limit is actually set by the observational limits that exist for Ba \citep{jacobson15}.

By contrast, the lower limit would be set by the regions of the collapsing star that were
able to undergo at least a partial r-process, but that produced a maximum amount of Ba
together with a minimum amount of Sr. We have used the r-process calculations of
\cite{woo94} to try to estimate what the extreme values of these two isotopes
would be. That paper studies the processing in a neutrino wind from the nascent neutron
star of a series of shells that are emitted from the surface of the star's core. The shells that
produce the values that are critical to our conclusions are also the ones that were found in
that study to produce a successful r-process, that is, they produced both the A = 130
and 195 peaks in reasonable agreement with those observed. 

In the study of \cite{woo94}, a succession of 40 trajectories
(i.e., thin shell wind elements), all originating deep within the
(spherically symmetric) star, but having different initial density,
temperature, entropy, and electron fraction, were emitted from the
star, thus contributing to the total r-process nucleosynthesis. The
bubble evolved in time, so that the conditions under which the
individual trajectories were processed changed with the identity of the
trajectory. We also assumed that the different trajectories were emitted
from the star successively, but ceased when the collapse to the 
BH occurred. This would be consistent with \cite{woo94}, who
assumed successive emissions of the trajectories to generate a good
representation of the solar r-process abundances.
\section{Model Calculations}
\label{calculations}
Code calculations are described in \cite{bfak} and are 
mentioned here for convenience.

The network calculation employed the {\it libnucnet} libraries
for the reaction networks \citep{meyer07} along with the nuclear and reaction data from the JINA reaclib database \citep{cyburt10}. As with 
previous calculations using this model \citep{bfak}, the
hydrodynamic model of \cite{woo94} was used as a starting point;
trajectories 24-40 were found in that study to produce a successful or partial r-process.
Network calculations began at the point where $T_9$ dropped below 2.5.
and used initial abundances from
the \cite{woo94} results. As with prior work, we assumed an initial abundance of massive nuclei from a single nuclear
seed (plus neutrons, protons, and $\alpha$-particles) derived from the average mass and electron fractions, $Y_e$ in the trajectories
of \cite{woo94}, corresponding to the result that the isotopic 
distribution is peaked around this seed.  Each calculation was continued
until the abundance distribution versus mass number had frozen out.

\begin{figure}
	\includegraphics[width=\linewidth]{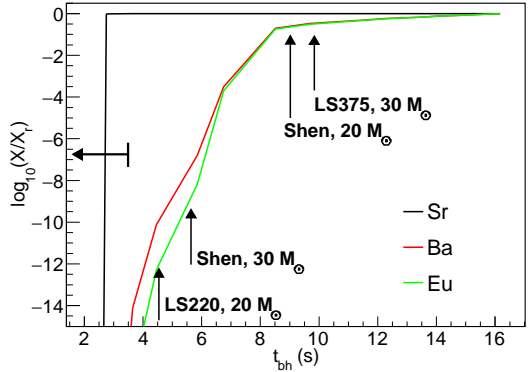}
	\caption{Sr, Ba, and Eu mass fraction in an r-process site as a function of
		the post-bounce BH collapse time in an r-process site.  The vertical
		arrows indicate combinations of Z=0, M$_{ZAMS}$ and EOS with collapse times greater 
		than 3.5 s.  The horizontal arrow indicates the range of collapse 
		times most prevalent in this study. 
		\label{fig3}} 
	%\label{fig3}}
\end{figure}

The Sr, Ba, and Eu mass fractions normalized to those of
a complete r-process are shown as a function of post-bounce collapse time
in Figure \ref{fig3} for various assumptions about the nuclear EOS \citep{bfak}, which 
contributes to the collapse time (discussed below).  
These data use the thermodynamic results 
from the late-time ejected
trajectories in the wind
model of \cite{woo94} and a network calculation to compute
ejected mass fractions as a function of post-bounce time. A value of one indicates that the resultant
mass fraction in the r-process site is equal to that of a complete r-process
(i.e., one in which all mass shells in the SN progenitor have been successfully ejected, in which an
ensuing r-process produces a robust A$\sim$195 peak).  In this explosion model,
the r-process does not proceed until after the ejected mass shell temperature drops well below $T_9$=10.  
This does not occur until after the mass shell is rapidly blown off the 
surface of the nascent neutron star.  The temporal evolution of Sr, Ba, and Eu production
depends not only on the progress of the r-process within that mass shell, but also on the
point in time at which the mass shell is ejected.  Sr is produced very early  
in the r-process evolution because the mass shell in which Sr is produced is ejected early on; 
only extremely early collapse times 
result in a significant reduction of the Sr production, while
the Ba and Eu abundances are built up more slowly in time; delayed BH collapse 
or failed explosions
can therefore cut off the later ejected mass shells, which produce more robust Ba and Eu
abundances.  Thus, any shells which may not be ejected due to fallback or a delayed failed
explosion will not contribute to the final r-process abundances, resulting in a partial
r-process abundance distribution. We use this model to parametrize the Sr, Ba, and Eu production as 
a function of collapse time in all collapse scenarios. 

\begin{figure}
	\includegraphics[width=\linewidth]{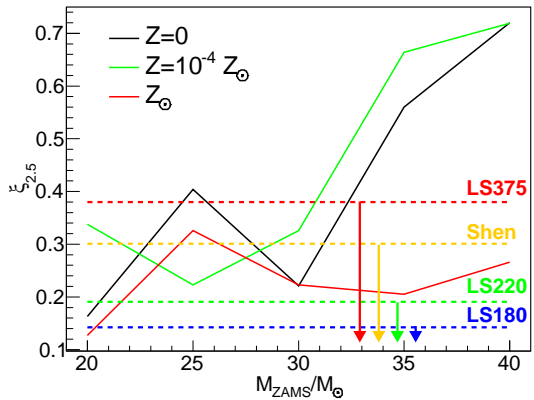}
	\caption{Relationship between the bounce compactness parameter $\xi$ and 
		the progenitor mass and metallicity \citep{GRID_ref}.  The horizontal lines
		indicate the values of $\xi$ for various EOSs for which the collapse time is 3.5 s.  Points
		below these lines have $t_{bh}>$3.5 s. The arrows indicate the range of $\xi_{2.5}$ for which the collapse time exceeds 3.5 s.
		\label{fig4}}
	%\label{fig4}}
\end{figure}

The relationship
between the stellar progenitor and the compactness $\xi_{M/M_\odot}$ has been shown in
previous works \citep{GRID_ref}, 
where the compactness parameter is defined at bounce as
\begin{equation}
\xi_{M/M_\odot} = \frac{M/M_\odot}{R(M=M_\odot)/\mbox{1000 km}}.
\end{equation}
In this case, it is convenient to choose the compactness at $M=2.5M_\odot$,
$\xi\equiv\xi_{2.5}$.  This relationship is shown in Figure \ref{fig4}
for various progenitor masses and metallicities, and varies slightly
with the nuclear EOS.  In this figure, progenitor masses are taken from
the compilation of \cite{ww02}.

\begin{figure}
	\includegraphics[width=\linewidth]{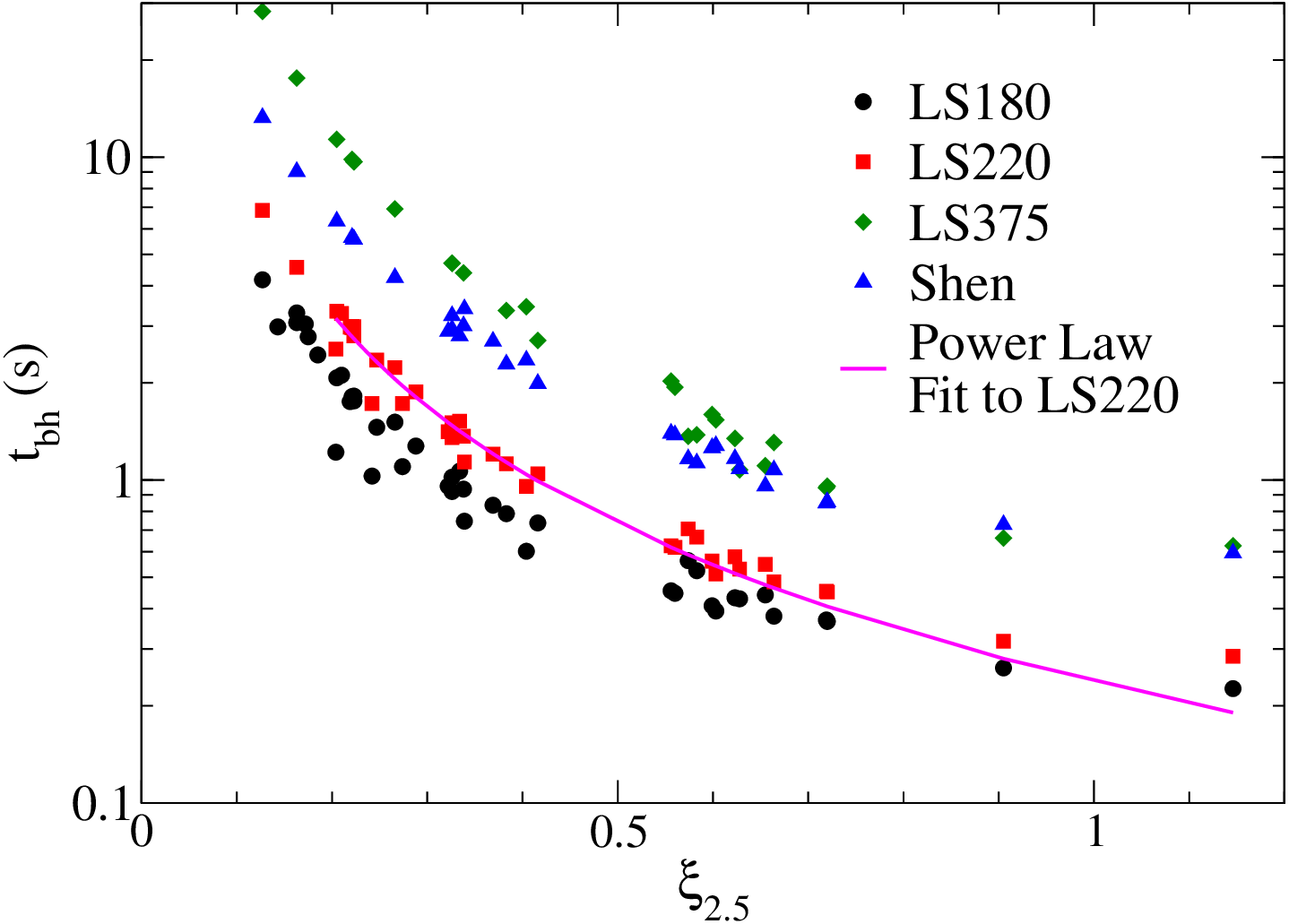}
	\caption{Relationship between the bounce compactness parameter $\xi$ and 
		the BH collapse time for several nuclear equations of state \citep{GRID_ref}.  
		The
		line corresponds to the $t(\xi)$ fit to the LS220 EOS.
		\label{fig5}}
	%\label{fig5}}
\end{figure}

The relationship between the compactness and the BH collapse time for
several nuclear EOSs is shown in Figure
\ref{fig5} \citep{oconnor11}.  Each data point represents the results of a GR1D calculation
for a particular progenitor model.  It can be seen that the collapse time
can be determined using only the compactness and the EOS.  In this way,
the details of the bounce, which depend on the stellar mass and metallicity,
are conveniently contained within the compactness parameter.  For each
EOS, the collapse time in seconds can be parametrized as
\begin{equation}
\label{t_vs_xi}
t_{bh} = A\xi^{-B}.
\end{equation}
In Figure \ref{fig5}, long collapse times (t$_{bh}>$3.5 s) are extrapoloated based on the value of $\xi$ in the collapse.  
The uncertainty
of these long collapses will be discussed later.
At the suprasaturation densities of neutron star interiors, a stiffer EOS,
such as the LS375, will result in a larger neutron partial pressure, and
the collapse time will be longer than in the case of a softer EOS, such as
the LS220.  

The collapse times in this figure can be
compared to those determined in prior evaluations \citep{sumiyoshi08}.  While they are comparable in models for
40 $M_\odot$ evaluated by \citet{sumiyoshi08},
the slightly longer collapse times in that work are 
most likely due to the fact that the progenitor models
used \citep{ww95} did not include mass loss resulting
in a higher compactness at bounce.

Given the results in Figures \ref{fig3}, \ref{fig4}, and \ref{fig5},
the Sr, Ba, and Eu yields
in a tr-process (normalized to those produced in a full r-process) 
can be determined as a function
of progenitor mass and metallicity.  As noted, these yields are minimum yields.  That
is, in the non-rotating spherical models employed here, the collapse times
in Figure \ref{fig5} are minimum collapse times.

It is noted, however, from Figure \ref{fig5} that the collapse time
as a function of $\xi$ becomes very sensitive to $\xi$ below a certain cutoff in $\xi$.  
Previous models have assumed that a collapse time $t_{bh}>3.5$ s represents a 
successful explosion \citep{oconnor11}. Similarly, other models have evaluated the
success rate of the explosion as a function of $\xi_{2.5}$ \citep{horiuchi14}.  In either case, the
explosion success is determined as a fraction of models calculated with the same zero-age main sequence
(ZAMS) mass, with model variations
determined by various model parameters for individual 
ZAMS mass.  
In the parametric model here,
	we evaluate the tr-process cutoff times in
	two different ways.  First, we assume
	that t$_{bh}>$3.5 s is possible, and evaluate
	these times in the tr-process model.  
	Second, we assume that any collapse longer
	than 3.5 s is not possible, and a full
	r-process results. This assumption, though arbitrary, is supported by prior work \citep{oconnor11}. Further, this allows us to gauge
	the uncertainty in the collapse in our model; if long collapse times are not possible, we have an estimate of the effect of not including these times in the calculation.

It is worthwhile at this point to mention the relationship between the collapse time, the
compressibility, and how a collapse or failed explosion is determined.  For example, let us consider
Figure \ref{fig3}.  Because the Sr-producing shells are ejected early on, as mentioned previously, 
Sr is produced earlier
than Ba or Eu, which is produced in shells ejected later.  Thus, any collapse which occurs in the 
intervening time between the Sr-producing shells and the Ba-producing shells will result in Sr-enriched
matter ejected into the ISM. Any collapse occurring after the ejection of the Ba-producing shells
will result in a more robust r-process.  Any collapse prior to this will prevent these Ba-producing
shells from begin ejected.  As it turns out, many of the stars in our
calculations have collapse times less than 3.5 s. This time range is shown as a horizontal
double arrow in Figure \ref{fig3}.
Examining this time range, it is noted that nearly all collapse scenarios will result in no production of Ba and Eu, while it
may result in partial or full production of Sr because the collapse time is less than the time
necessary to produce Ba.  The tr-process is binary in this respect.

There are then three regions of interest in Figure \ref{fig3}.  The
	first is for collapse times between about 3 and 4 s.  For these stars, 100\% of the produced
	Sr is ejected, and essentially no Ba and Eu is ejected.   These would include the more massive
	and some intermediate-mass stars, depending on the EOS.  The second region is the region where
	the collapse time is less than 3 s.  In this region, the Sr abundance drops precipitously.  
	This region would include the most massive stars in this model.  

The third region is perhaps the most interesting.  This is the
	region where the collapse time is greater than about 4 s.  Using Equation \ref{t_vs_xi} and 
	Figures  \ref{t_vs_xi} and \ref{fig4}, collapse times can be determined for various
	masses and EOSs.  Stars with calculated collapse times between 4 s and 17 s are shown 
	in this figure.  For these stars, this model assumes that - while 100\% of the produced Sr is 
	ejected into the ISM - only a fraction of the Ba and Eu is ejected.   
	These stars are assumed to have successful explosions.  However, in this model, consistent with the
	assumptions of the tr-process, we assume this computed collapse time in creating a parameter to
	gauge the amount of ejected material.  This also helps to compensate for the binary nature of
	our model in terms of explosion or failed explosion.  That is, the fraction of failed explosions for
	stars in the intermediate mass range \citep{horiuchi14, oconnor11} is simulated in our model by simulating a fractional r-process for stars in this range.

We can also obtain a rough idea of whether or not a star will collapse in this model and how.  Using this result and using the cutoff time of 3.5 s as the time above which no Ba and Eu is
	produced (also the time above which no collapse
	is assumed to occur in the referenced collapse model \citep{oconnor11}), one can determine the compactness for which a star will then collapse using 
	Equation \ref{t_vs_xi}. This critical compactness, $\xi_{c}$, depends on the EOS, and is shown in Figure 
	\ref{fig4}.  For a chosen EOS, any star with $\xi_{2.5}>\xi_{c}$ will 
	collapse at a time greater than 3.5 s.  For example,
	for a stiffer EOS in this model, only the most massive low-metallicity stars will result in a collapse at $t_{bh}>$3.5 s.  However,
	for the softest EOS, all but the least massive stars will result in a collapse.  Because of the binary 
	nature of this model (collapse probability is either zero or unity), and because a collapse
	generally (but not always) results in only Sr escaping the star, this becomes relevant in
	relating the [Sr/Ba] ratio to the predicted GCE abundances.  If only the most massive stars collapse
	as in the case of a stiffer EOS, one expects lower overall [Sr/Ba] values, whereas if only the
	lighter stars successfully explode, one expects higher overall [Sr/Ba] values.  Because the collapse
	probability is set to unity in this model, an upper limit on [Sr/Ba] can be placed in the GCE 
	calculation.
\section{Results}
\label{results}
\begin{figure*}
	\includegraphics[width=\textwidth]{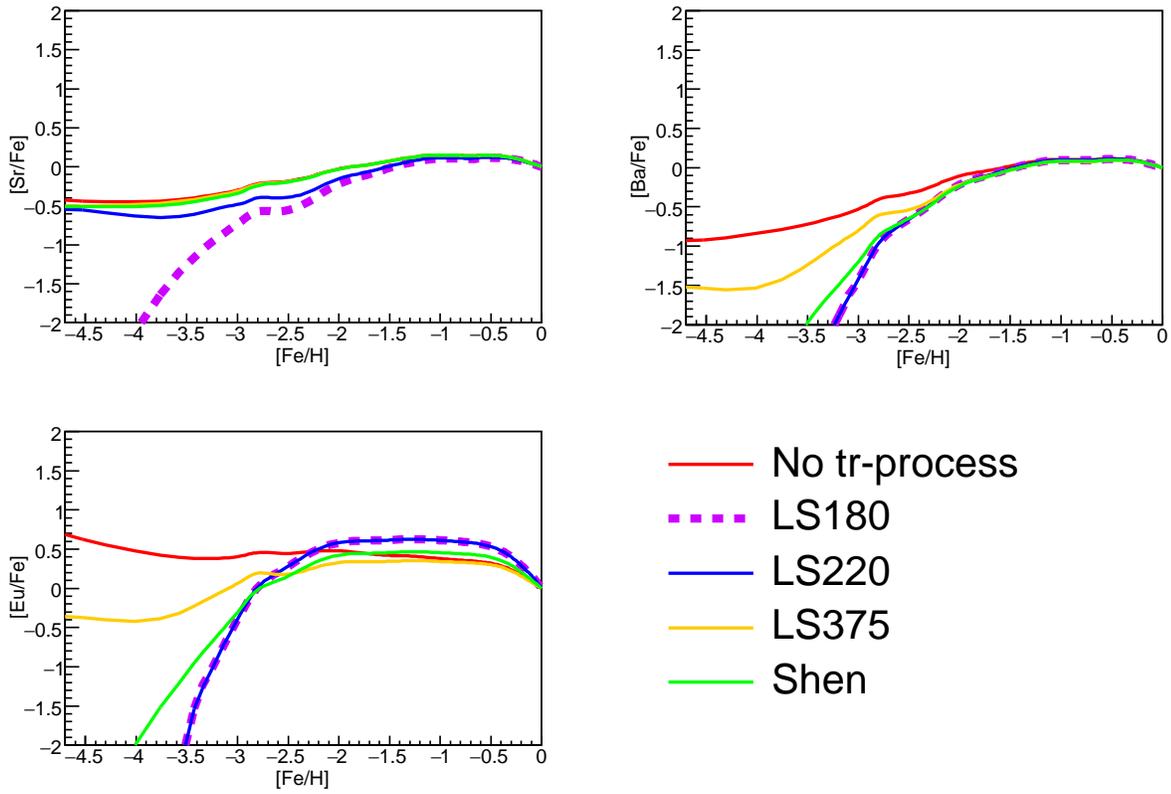}
	\caption{GCE results for [Sr/Fe], [Ba/Fe], and [Eu/Fe] 
		as a function of metallicity assuming a tr-process
		with various assumptions of the nuclear EOS.  For the [Ba/Fe] and [Eu/Fe] values,
		the lines for the LS220 and LS180 EOSs are on top of each other.
		\label{fig6}}
	%\label{fig6}}
\end{figure*}

Assuming a tr-process model resulting in Sr, Ba, and Eu production yields given
in Figure \ref{fig3}, the [Sr/Fe], [Ba/Fe], and [Eu/Fe] values as a function
of metallicity [Fe/H] are shown in Figure \ref{fig6} for various assumptions 
of the EOS.  Four different EOSs are examined in this calculation.  These
include three variations of the Lattimer-Swesty EOS \citep{lseos} and the Shen EOS \citep{shenn}.  
It is clear that the EOS
has a significant effect on the value of [Ba/Fe] and [Eu/Fe] as a function of stellar metallicity
with only a slight effect on the [Sr/Fe] values.
A tr-process event will produce less Ba and Eu for
a softer EOS.  
Roughly correlating metallicity, [Fe/H] to galactic age \citep{timmes}, we can explain this
behavior as follows. 

A stiffer EOS will have a longer collapse time as the 
neutron partial pressures are larger in the high-density, isospin-asymmetric core.  This
longer collapse time can permit more material ejection during the r-processing phase. The 
net result is that the r-process progenitor element ejection into the ISM will not vary as much from that
of a typical r-process (in which the ``collapse time,'' $t_{bh}$ is assumed to be infinite).  On the
other hand, for a softer EOS, the faster collapse time will significantly reduce the amount of
Ba and Eu ejected as this is produced in layers closer to the core, which are ejected later.
For an earlier collapse time, the cutoff in production will occur prior
to the ejection of the mass shells responsible for the production of r-process elements beyond
the A$\sim$130 peak.  Since the collapse time also roughly depends on progenitor mass, more massive
stars will produce higher [Sr/Ba] ejecta earlier in galactic history, shifting [Sr/Ba]
as a function of [Fe/H] in GCE predictions.

The effects of the collapse on Sr production are not as pronounced, since Sr is produced in
the very early stages of the r-process (and likely has a contribution from the s-process as 
later times in the galactic evolution corresponding to higher metallicitiy).  In fact, only very 
early collapse times corresponding to stars with progenitor masses M$_{ZAMS}\gtrsim$25M$_\odot$ will 
result in a tr-process in which the Sr yields are significantly reduced.
Given the Salpeter initial mass function (IMF) used in this model, this would comprise a small fraction of the Galactic Sr
yield.  

\begin{figure}
	\includegraphics[width=\linewidth]{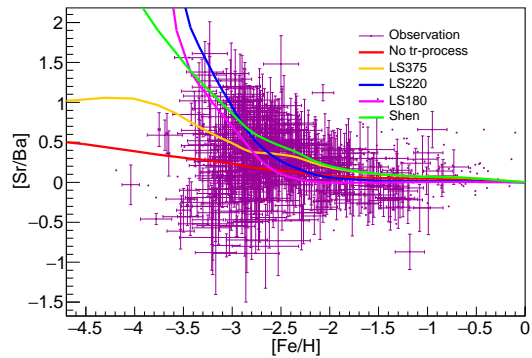}
	\caption{Results of the GCE calculation showing [Sr/Ba] as a function of
		metallicity halo stars produced in a tr-process for various 
		assumptions of the nuclear EOS and collapse. These are for no tr-process,
		and several additional EOSs. The assumption of an
		LS375 EOS closely resembles the case of no tr-process. The stars
		potentially affected by the s-process shown in Figure 
		\ref{fig1} are excluded in this figure.
		\label{fig7}}
	%\label{fig7}}
\end{figure}

\begin{figure}
	\includegraphics[width=\linewidth]{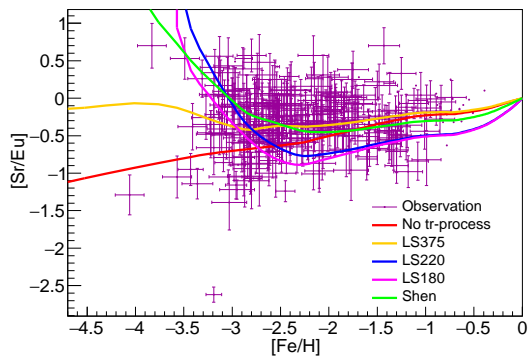}
	\caption{The same as Figure \ref{fig7} but for [Sr/Eu].
		\label{fig8}}
	%\label{fig8}}
\end{figure}

The calculated [Sr/Ba] and [Sr/Eu] values are also shown in 
Figures \ref{fig7} and \ref{fig8} compared to observed values.
The significant changes in the Ba and Eu ejection in a collapse scenario results in a dramatic
change in the [Sr/Ba] and [Sr/Eu] ratios.  
In a previous paper \citep{aoki_apj}, changes in
[Sr/Ba] were suggested to be caused at least in part by turbulent ejection of material in a
collapse scenario.  It is seen that for a softer EOS, the minimum(maximum) values in 
[Ba/Fe]([Sr/Ba]) and  [Eu/Fe]([Sr/Eu]) as a function of metallicity can be achieved in a tr-process
for partial enrichment of r-process elements in a GCE model.  As noted above, the GCE results shown in these figures
represent extremes in these ratios as they are produced in collapse scenarios corresponding to the minimum
collapse time to a BH.  

In examining the abundance ratios of [Sr/Ba] and
[Sr/Eu] as they relate to the EOS, one sees that these ratios generally increase as the EOS
softens.  However, at some point, the EOS becomes so soft that the collapse time becomes early enough
to prohibit Sr ejection, and the ratios of [Sr/Ba] and [Sr/Eu] begin to decrease with the softness of the EOS.
This may occur for an EOS with a softness somewhere between the Shen EOS and the LS220 EOS,
as one sees that the [Sr/Ba] and [Sr/Eu] ratios calculated using an LS220 EOS drop below those 
calculated using a Shen EOS at metallicities -2.5$<$[Fe/H]$<$-2.

We also note the interesting structure of the [Sr/Ba] values for the Shen EOS. 
	While we don't place a large quantitative weight on this model, we point out that the
	Shen EOS is not quite as stiff as the LS375, but 
	much stiffer than the LS220 EOS.  For the Shen EOS, 
	20 and 30 $M_\odot$ Z=0 stars collapse (see Figure \ref{fig4}), but with long
	collapse times.  Also, 25 $M_\odot$ stars with Z=0 and Z=Z$_\odot$ explode with collapse times less than 3.5 s, but 25 $M_\odot$ stars with Z=10$^{-4}Z_\odot$ do not. These stars may be major
	contributors to Ba enrichment at [Fe/H]$\sim$-3.

One sees from an examination of Figures \ref{fig6} - \ref{fig8} that
there appears to be a sharp lower limit in the [Ba/Fe] and [Eu/Fe] ratios as a function of
metallicity corresponding to a sharp upper limit in [Sr/Ba] and [Sr/Eu].  We note that a softer
EOS will produce this trend.  However, if the EOS is not soft enough, this trend cannot
be produced by a tr-process alone.
On the other hand, if the EOS is too soft, the lower(upper) limit in [Ba/Fe]([Sr/Ba]) and [Eu/Fe]([Sr/Eu])
exceeds the limits of the observations.  These observed limits then provide observational constraints on
the stiffness of the nuclear EOS.  In comparing these results to those of \cite{aoki_apj}, one may
conclude that it is possible that the tr-process may be responsible for not only the upper and lower
limits in the heavy metal production early in the galaxy, but it may also be responsible for the observed
scatter in these elemental abundances.  The extremes are directly dependent on the EOS in this model.

We also note that in all cases, the value of [Sr/Ba]$>$[Sr/Eu], regardless of the EOS chosen or the collapse time cutoff chosen.  This is consistent with observations (see Figures \ref{fig7} and \ref{fig8}).  
This results primarily from the fact that Eu is produced
in later ejecta in the explosion model.  Since it is ejected
at later times, the overall value of [Sr/Ba] is expected
to exceed that of [Sr/Eu].
\subsection{Uncertainties in Results}
\begin{figure*}
	\includegraphics[width=\textwidth]{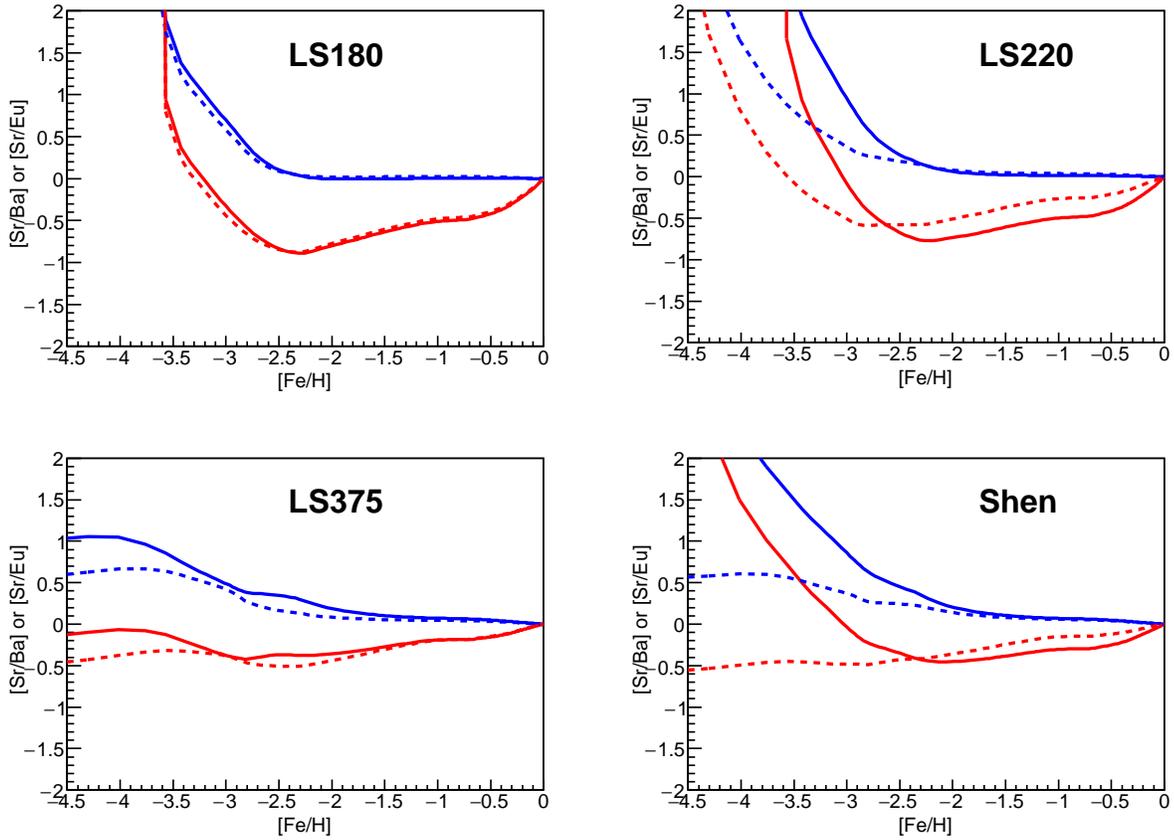}
	\caption{GCE results for [Sr/Ba] (blue lines) and [Sr/Eu] (red lines)as a function of metallicity for the four 
		EOSs studied.  The solid lines are results assuming arbitrary collapse times.  The dotted lines assume that 
		collapse times are constrained to be below 3.5 s.
		\label{fig9}}
	%		\label{fig9}}
\end{figure*}

	In Figures \ref{fig7} and \ref{fig8}, an arbitrarily long collapse time (as a function of $\xi$)
	has been assumed.  However, these longer collapse times may be uncertain, so 
	we evaluate the uncertainty induced
	by assuming a collapse time.
	To do this, we repeat the calculation of
	Figures \ref{fig7} and \ref{fig8}, but we assume that any collapse time in excess of 3.5 s is not possible, and the explosion mechanism produces a complete r-process (i.e., no tr-process). Exclusion of these longer collapse times from the tr-process, while somewhat arbitrary, corresponds to prior evaluations, 
	in which the calculated collapse time is assumed to be 
	less than 3.5 s \citep{oconnor11}.
	The results of this analysis are shown in Figure \ref{fig9}.  In this figure, we compare the GCE calculations of Figures \ref{fig7} and \ref{fig8} to results of the same GCE evaluation; however, we assume that collapse times cannot be greater than 3.5 s.  That is, if a collapse time is determined to be greater than 3.5 s,
	then a full r-process results in that scenario since all
	shells in the SN are ejected.  In these figures, the range between the dashed and solid lines is the range for which this model can be taken to predict the [Sr/Ba] and [Sr/Eu] values.  
	This analysis is applicable to stars with collapse times between 3.5 s and
	$\sim$16 s (for which a full r-process will occur in the ejecta).  Collapse times longer than 	16 s 
	will not be affected (even if the stars collapse at this time) because
	the tr-process will have already produced  Ba and Eu abundances matching those of the r-process.

	We see from Figure \ref{fig9} that the very soft LS180 EOS does not change significantly if long collapse times are neglected.  This is because for the LS180 EOS, only the least massive stars escape collapse into a BH. We can expect this, if we examine
	Figure \ref{fig4}. In this figure, nearly all stars fall above the 3.5 s limiting line for the LS180 EOS, indicating that nearly all stars will collapse above 20 M$_\odot$, and no r-process is possible.  

	For the LS220 EOS, 
	however, it would appear that if we 
	assume a full r-process for a calculated collapse time greater than 3.5 s, then Ba and Eu is produced earlier on in galactic history.  This is because Z=0, 20 M$_\odot$ stars, with calculated collapse
	times of just over 4.5 s, are now able to 
	eject Ba in an r-process event. (These 
	stars are indicated in Figure \ref{fig3}.)  If a collapse time of $\sim$4.5 s is assumed, then a negligible amount of Ba and Eu is produced.
	Given the larger population of 20 M$_\odot$ stars in the IMF, this shifts the [Sr/Ba] drop to earlier times in galactic history.  
	Most of the collapse times are in a range
		between about 1 and 3.5 s in this work, as indicated in Figure \ref{fig3}.  This range is
		important because for collapse times in this range, either all
		or none of the Sr will be produced.  However, it
		can be seen from this figure that, in general (though not always), a collapse will result in a maximum amount of
		Sr ejection but no Ba ejection.
	
	A small change is noticed for the
	very stiff LS375 EOS. Very little change is expected because nearly all stars explode successfully (or have collapse times long enough to produce a full r-process).  This model shifts slightly closer to the model with no tr-process.
	
	Perhaps the most dramatic shift is for the the Shen EOS.  If a full r-process is assumed for long collapses, then the Shen assumption
	produces results which approach those of no tr-process.  In the case of the
	Shen EOS, several stars have collapse times greater than 3.5 s, but less than 16 s.  These are the Z=0, 20 M$_\odot$ and 30 M$_\odot$ stars,  the low-metallicity 25 M$_\odot$ star,
	and all but the 25 M$_\odot$ solar metallicity star.  Moving production of
	these collapse scenarios from a tr-process to a full r-process can dramatically change the outcome.

If this production is a result of the tr-process, then the 
EOS
dependence of the collapse time has a direct effect on the Sr and Ba
production.  Since the plotted values of [Sr/Ba] are 
maximum values, then it would appear that in this model  
a very stiff EOS cannot produce the upward scatter in [Sr/Ba].  A
softer EOS is necessary.  It's also noted that the upper limit of
observed [Sr/Ba] values provides an observed lower limit on the
softness of the EOS.
\subsection{Turbulence and Mixing}
Because of the 1D nature of the collapse
studied and the phenomenological assumption that a collapse or fallback
is directly related to shell ejection time of a shell in a SNII, mixing and
turbulent ejection is not dealt with in a strict quantitative fashion.
However, a simplistic evaluation of how mixing and turbulence may affect fallback and collapse scenarios is provided.

\begin{figure}
	\includegraphics[width=\linewidth]{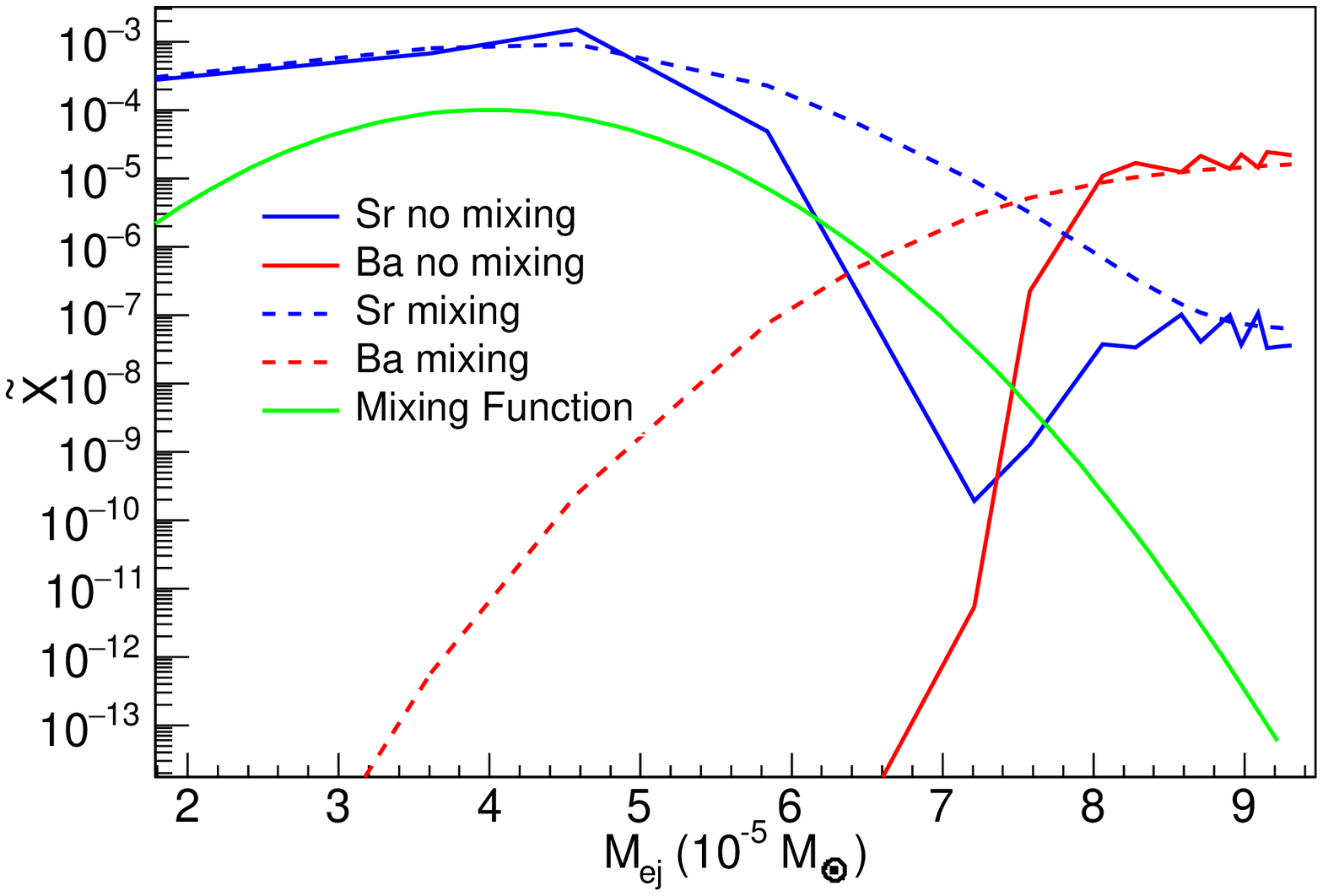}
	\caption{Example of differential normalized Sr and Ba yields as a function of ejected mass
		with and without turbulent mixing.  Also shown is a sample Gaussian mixing function $\omega(m-m^\prime)$ centered at 
		4$\times$10$^{-5}$ M$_\odot$ with $\sigma$ = 8$\times$10$^{-5}$.
		\label{fig10}}
	%		\label{fig10}}
\end{figure}

Two situations
are studied.  For a tr-process, the ejected mass fraction $X_i$ of an element ejected
from the proto-neutron star is a sum of mass fractions for individual mass elements up to the mass cutoff in the collapse
\begin{equation}
\label{yeild_no_turb}
X_{ej} = \frac{1} {M_{cut}} \int_{M_{cut}}^{M_{tot}}X(m)dm,
\end{equation}
where $X(m)$ is the amount ejected for a shell of mass
$dm$ at mass $m$ and the mass cutoff is determined by the collapse time.  The point below which material falls back to
the surface of the neutron star is at $M_{cut}$, and 
$M_{tot}$ is the outer-most mass shell in the SN in which 
an r-process occurs.
In the case of turbulent mixing, the mass fraction of a species
as a function of the ejected mass can be mixed between mass shells and is represented
by a convolution between individual shells.  The adjusted mass fraction $\tilde{X}(m)$ is:
\begin{equation}
\label{turb_mix}
\tilde{X}(m) = \frac{1}{M_+-M_-}\int_{M_-}^{M_{+}}X(m^\prime)\omega(m-m^\prime)dm^\prime
\end{equation}
where $M_+$ and $M_-$ is the top and bottom mass shell of the mixing zone, or the maximum and minimum of the mixing function, respectively. 
The function $\omega(m-m^\prime)$ is a function representing mixing between a shell at mass $m$ and one at mass $m^\prime$.  This function has a peak at $m=m^\prime$ and decreases monotonically for values above and below $m$. 
In this model, $\omega(m-m^\prime)$ is negligibly small at
$M_-$ and $M_+$
The mixing is taken over all shells, including
those below the mass cutoff.  In this case, a mixing of matter between adjacent shells is implemented.  For rapid
changes in yields between adjacent shells, this can have large effects around the mass cut.  We refer to this 
as the ``turbulent mixing'' case. For this example, the weighting function is taken to be a Gaussian function of
width $\sigma_M$=5$\times$10$^{-6}$M$_\odot$.
The net result is that isotopic composition in an ejected shell becomes a weighted average over that shell and adjacent shells.  In general, this should have little overall effect for a complete explosion in this model.  However, for a partial explosion, matter below the mass cutoff may be moved outward in 
the ejection.
The concept of turbulent ejection is shown in Figure \ref{fig10}, which shows the differential normalized yields of 
Sr and Ba as a function of mass coordinate of the ejected shell. The net effect is not to change the overall amount of material that is
ejected in an r-process, but to redistribute it somewhat.  In this case, there is little effect on the Sr in a tr-process
as most of that is ejected early on, and only short collapse times will reduce the Sr abundance. One sees
that the Ba ejection as a function of mass is increased by several orders of magnitude for early stage (low mass)
ejecta.

Another effect of turbulence, referred to as the ``turbulent ejection'' case, involves the preferential
ejection of matter into the ISM due to ejection of pockets of material near the collapse radius.  This is
similar to the turbulent mixing case except that the turbulence is assumed to occur in a specific location
of the star, thus, the weighting function is fixed about a specific mass element $m_o$ in the neutrino-driven wind
\begin{equation}
\label{turb_ej}
{X}(m) =\frac{1}{M_+-M_-}\int_{M_-}^{M_{+}}X(m^\prime)\omega(m-m_o)dm^\prime .
\end{equation}
In this case, only mass elements near $m_\circ$ are ejected.
Turbulent ejection is very similar to turbulent mixing in that the amount of matter does not change in a complete r-process.
However, it is redistributed by weighting one particular mass shell over the other. One may think of this as moving material from
one mass shell to another, creating a ``turnover'' or "exchange" in the mass distribution.  The treatment of the differential ejecta is
very similar to that shown in Figure \ref{fig10}, but a single mass element from a single area is redistributed.  In this
case, if the mixing is near the early stage ejecta, Sr near the surface of the star is mixed with deeper layers, resulting in
a somewhat delayed ejection of Sr.  The ramifications for a tr-process are clear in that the Sr ejection may be reduced 
for slightly later stage collapse times.  The redistribution of
mass in this model may result in a reduced amount of Sr ejected early on.  The earlier stage collapse times will then
result in a reduction in the Sr production while the Ba production, with the bulk of its yield ejected later, remains largely
unchanged. 

The turbulent mixing case results in a sort of ``averaging'' between 
individual mass elements in the ejection.
This is explained physically by mixing between adjacent shells.  In this case, 
some matter which would otherwise
be ejected then falls below the mass cut in the collapse.  On the other hand, some
material which would otherwise fall below the mass cut in the collapse is then ejected above
the cut due to mixing and escapes the star.  In this case, Ba production is more strongly affected
as it is produced closer to the protoneutron star surface.  For ejection times 
corresponding to mass cuts at smaller radii, Ba which is ejected much later in the SN
explosion is mixed near the mass cut.  The averaging of ejecta producing significant amounts
of Ba with adjacent shells producing only a small amount of Ba can result in small changes in
the Ba production around the mass cut.  However, this type of mixing has significant effects on the tr-process
for mass cuts that occur between shells producing significantly different amounts of Ba.

In the simple turbulent ejection assumption here, the temporal characteristics of the ejection are computationally outside
of the scope of this work.  However, two possibilities can be examined.  
One is that the mass shells emphasized in Equation \ref{turb_ej} are ejected at times corresponding to their ejection
times in Figure \ref{fig3} maintaining the order in which material is ejected, but emphasizing ejection
of particular shells over another.  In another case, a turbulent ejection mechanism can be assumed in which 
a single shell is promptly ejected independent of any subsequent collapse. The latter case represents an extreme
in the ejection model and is useful for studying the scatter in Sr and Ba in single-site r-process considerations.

\begin{figure}
	\includegraphics[width=\linewidth]{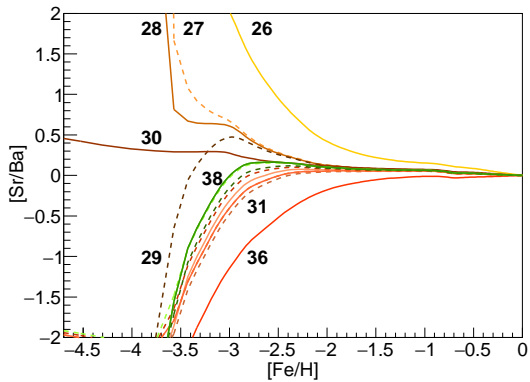}
	\caption{GCE results for different assumptions concerning turbulent ejection of individual 
		mass elements in this conceptual model.
		Results assuming ejection of individual mass shells are shown.
		That is, each line in the figure is the [Sr/Ba] assuming that
		only a single shell is ejected (normalized to the 20 M$_\odot$ of \cite{woo94}) in every explosion.  
		The numbers indicate which trajectory in the Woosley (1994) model is ejected.  
		The LS220 EOS is assumed with a mixing function $\omega(m-m_\circ) = \delta(m-m_\circ)$ where $m_\circ$ is an individual mass shell.
		\label{fig11}}
\end{figure}

For the case of turbulent ejection, preferential emission of individual shells is assumed to occur for
stars with mass greater than 20 $M_\odot$.  By asuming a single shell emission, the yield values for a 
tr-process GCE calculation are then scaled to the values of the Sr and Ba yield fractions for a particular shell.
The results of the GCE calculation assuming a model in which only single shells are emitted via a turbulent
mechanism are shown in Figure \ref{fig11}.  This calculation assumes that production in an r-proces site 
is due to the ejection of only a single indicated shell in the figure.  The weighting factor 
$\omega(m-m_o) =  \delta(m-m_o)$, which corresponds to mixing
between two shells, but not necessarily adjacent shells.  This ejection proceeds independently of the subsequent collapse.  Note
that while the trend favors Sr production near the surface and Ba production near the core, the trend
is not necessarily monotonic with shell number for closely-packed shells near the core.  This seems 
consistent with non-monotonic environmental parameters (e.g., entropy) in the original model \citep{woo94}.
This effect may also be responsible for the non-monotonic nature of the [Sr/Ba] values with
EOS and [Fe/H] in Figure \ref{fig7}.

One notes the extremes in [Sr/Ba] in Figure \ref{fig11}.  For shells ejected near the surface, 
the final composition is nearly all Sr, while shells ejected deeper in the star produce nearly all
Ba.  The resultant range (or scatter) in [Sr/Ba] from this calcualtion covers almost exactly the 
entire range of observed [Sr/Ba] as a function of [Fe/H] in Figure \ref{fig1}, indicating that 
a turbulent ejection mechanism in a collapse scenario may be responsible for the large
variation in yields in a core collapse scenario.

\begin{figure}
	\includegraphics[width=\linewidth]{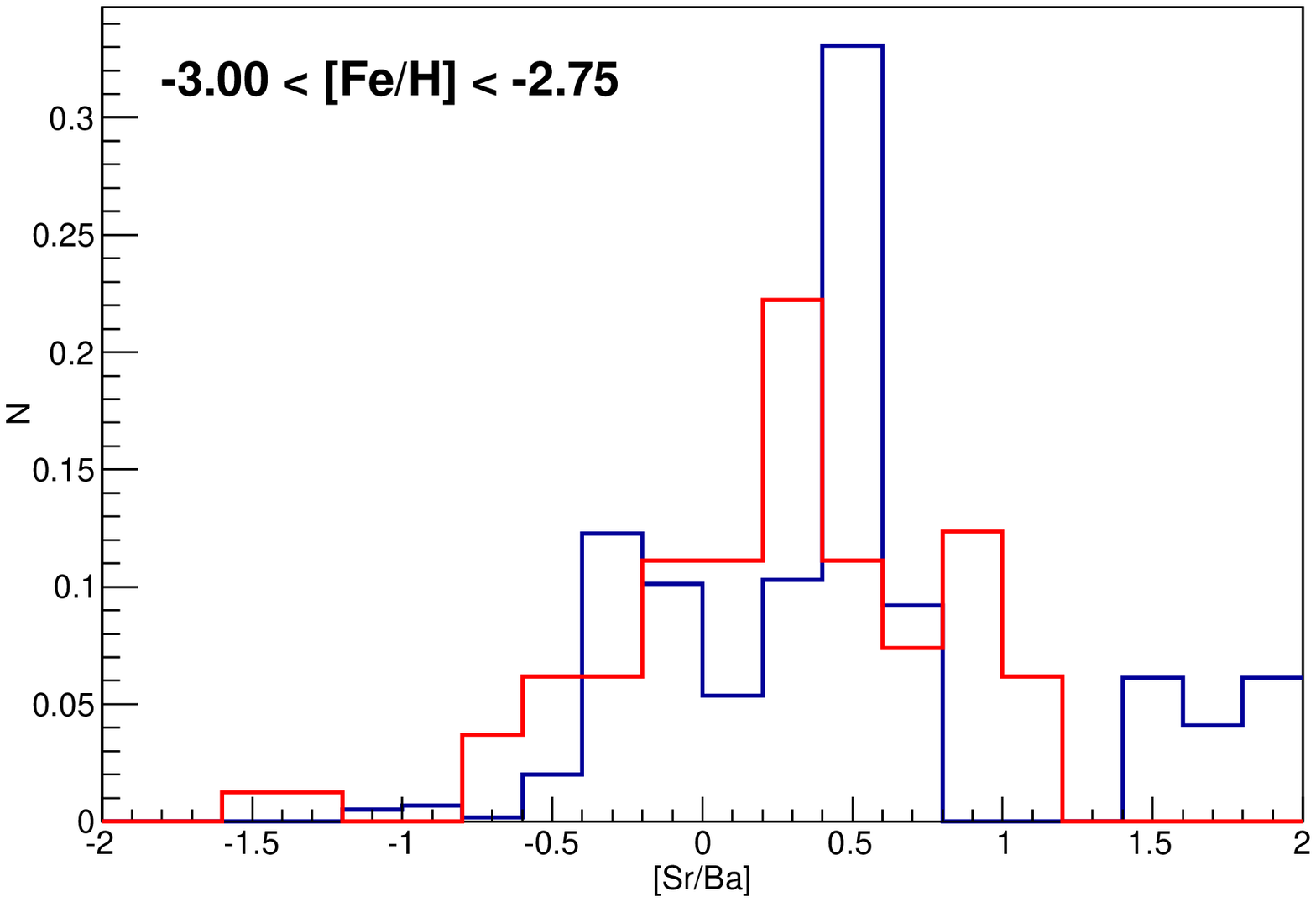}
	\caption{Mass weighted projection of [Sr/Ba] values for metallicity -3.0 $<$ [Fe/H] $<$ -2.75.  GCE results assuming single-shell ejection with
		an LS220 EOS (blue line) are compared to observational data (red line).
		\label{fig12}}
	%		\label{fig12}}
\end{figure}

If one makes the assumption that turbulent ejections can occur with equal probability for any of
the r-processing shells, we plot the mass-weighted distribution of [Sr/Ba] for
single-site r-process enriched stars in our model by projecting the plot 
in Figure \ref{fig11} for a specific range of [Fe/H].  An example of this is shown in 
Figure \ref{fig12} for -3.00 $<$ [Fe/H] $<$ -2.75 compared to the same projection
from the observations shown in Figure \ref{fig1} at the same range in 
metallicity.  Each [Sr/Ba] bin from the GCE results shown 
in the projection is weighted by the mass of the shell
producing that value of [Sr/Ba].  We see that, though the distributions
are similar (e.g., most of the stars at this metallicity have [Sr/Ba] $\sim$ 0.5,
there are some variations in the comparison.  These variations are not surprising 
as the turbulent ejection is not expected to be equally weighted for all mass shells.
That is, turbulence may favor deeper or more shallow mass shells in a true
core collapse explosion event.  For more elaborate models like those of numerical hydrodynamic calculations perhaps this
comparison can be used to constrain the amount of turbulence and weighting
of turbulence in an event.
\section{Conclusions}
\label{conclusions}
A model was presented in which the [Sr/Ba] values in 
EMPs are assumed to result from enrichment from partial explosions of r-process progenitor events in
the early galaxy.  A large scatter in [Sr/Ba] can result as BH
collapse in SNII will cause a truncation in the ejecta and the
r-process yields which depends heavily on the progenitor mass.  In this model, a maximum
in the [Sr/Ba] values as a function of metallicity is due to a minimum
collapse time in BH collapse.

The results presented here are compared to observations of EMPs.  A 
fascinating result is that, despite the large scatter in [Sr/Ba], no
observations exist above a sharp line corresponding to a maximum in
the values of [Sr/Ba] at specific metallicities.  It has been shown
that in the tr-process model, this value is related to the 
stiffness of the nuclear EOS. 

We stress that this model is used to explain the light-element r-process enrichment in metal-poor stars. 
However, this does not exclude production of heavy r-process element
enrichment in metal-poor stars.  This enrichment may be due to
asymmetric explosion mechanisms, high angular momentum explosions,
or other r-process sites such as neutron star mergers.

We note that the developed model may not be sophisticated enough to 
determine the nuclear EOS, but it does provide a qualitative suggestion towards
a softer EOS.  This is consistent with results from prior experimental work \citep{famiano06}.
Observations of
neutron star masses \citep{lattimer12,lattimer} then provide complementary
constraints.  However, most fascinating is the 
prospect that while neutron star masses may determine a minimum in
the stiffness of the nuclear EOS, observations of [Sr/Ba] as a function
of metallicity may provide a measure of a {\it maximum} in the stiffness of the
nuclear EOS.  That is, a maximum in the [Sr/Ba] as  function of metallicty
is an indicator that the nuclear EOS must be soft enough to produce such
an observation.

It is shown that the apparent scatter in [Sr/Ba] at fixed metallicity may result
from an inversion of
individual elements during the collapse and subsequent ejection of material from the 
surface of the proto-neutron star.  We have studied the extreme cases in which
only a single mass element is ejected prior to a possible collapse in massive
stars \citep{aoki14}.  For mass shells which produce significant amounts of Sr, very little 
Ba is produced and vice-versa.  This can result in a relative enhancement of a particular element possibly explaining the scatter in [Sr/Fe], [Ba/Fe],
and [Sr/Ba].  It is interesting that the scatter in our calculations closely matches
that of the observational data.

It's also interesting to note that [Sr/Ba]$>$[Sr/Eu] consistent
with observations, indicating that the tr-process preserves
relative abundances in GCE observations.  Further study and
observation to compare
tr-process results and observations is desired.

The resulting [Sr/Ba] evolution was found to depend strongly on the EOS.  Future
work will seek to improve the uncertainty in these results.  These
uncertainties include the assumption of a universal Sr and Ba
yield as a function of collapse time independent of the progenitor mass
and metallicity as well as the effects from neutrino luminosity
profile changes in the collapse.

The proposed model may be useful
in future studies in which r-process enrichment in metal-poor stars can be explained
with various mechanisms.  It is certainly worth expanding the current study to improve on
some of the approximations and assumptions in this model.  This would include a more
accurate treatment of time relationship of r-process ejecta as a function of progenitor
mass, an exploration of turbulent ejection in core collapse scenarios, a study of
the uncertainties in stellar collapse time as it relates to
the dynamics of the collapse, and more complex multi-dimensional collapse calculations.  
Also, the mechanism of ejection should
be carefully examined.  The model assumes a core-collapse scenario, though it is
applicable to an r-process in a neutrino-driven wind or in a magnetohydrodynamic jet as both
can experience a collapse to a BH.
In addition, while this proposed model assumes a dynamic effect of 
material accretion, the effects of neutrino luminosity changes in a collapse may
also be a significant cause of the tr-process.
\acknowledgments
MAF's work is suppoted by NSF grant \#PHY-1204486 and \#PHY-0855013;
TK's by Grants-in-Aid for Scientific Research of JSPS 
(26105517 and 24340060), of the Ministry of Education, Culture, Sports,
Science and Technology of Japan.
This work was supported through the NAOJ Visiting Professor Program. WA and TS were supported by the JSPS Grants-in-Aid for Scientific
Research (23224004).  The authors greatly appreciate input from
and discussion with Richard Boyd.

\clearpage

%\begin{table}
%\caption{GCE Parameters Used in This Model\label{gce_par}}
%\begin{tabular}{lcr}
%\hline
%{\bf Parameter} & {\bf Value} & {\bf Units}\\
%\hline
%\hline
%Schmidt star formation efficiency & 2.8      & Gyr$^{-1}$\\
%Schmidt star formation exponent   & 2.0      & ~ \\
%Salpeter exponent                 & -1.31    & ~ \\
%Lower limit of IMF                & 0.08     & M$_\odot$ \\
%Upper limit of IMF                & 40.0     & M$_\odot$ \\
%Distance to solar neighborhood    & 8.5$\times$10$^3$ & pc \\
%Solar vicinity surface density    & 75.1     & M$_\odot$/pc$^2$ \\
%Galactic center surface density   & 10$^4$   & M$_\odot$/pc$^2$ \\
%Extent of inverse square profile  & 2000.0   & pc \\
%Age of Galaxy                     & 15.0     & Gyr \\
%Timescale of disk formation       & 4.0      & Gyr \\
%Lower limit for Type II rates     & 11.0     & M$_\odot$ \\
%Upper limit for Type II rates     & 40.0     & M$_\odot$ \\
%Lower limit for binary systems    & 3.0      & M$_\odot$ \\
%Upper limit for binary systems    & 16.0     & M$_\odot$ \\
%\hline
%\end{tabular}
%\end{table}

\clearpage

\begin{thebibliography}{}
\bibitem[Aoki et al.(2007a)]{aoki07}Aoki, W. et al. 2005, ApJ, 632, 611
%
\bibitem[Aoki et al.(2007b)]{aoki07b}
Aoki W., Beers T.C., Christlieb N., Norris J.E., Ryan S.G., \& Tsangarides S. 2007, \apj, 655, 492
%
\bibitem[Aoki et al.(2013)]{aoki_apj}Aoki, W., Boyd, R.N., Kajino, T., \& Famiano, M.A. 2013 \apj, 766, L13
%
\bibitem[Aoki et al. (2014)]{aoki14}Aoki, W. at al. 2014,
Science, 345, 912
%
\bibitem[Arcones \& Montes (2011)]{arcones11}
Arcones, A. \& Montes, F. 2011, \apj, 731, 5
%
\bibitem[Asplund et al.(2009)]{asplund09}
Asplund, M., Grevesse, N., Sauval, A.J., \&
Scott, P. 2009, ARAA, 47, 481
%
\bibitem[Boyd et al.(2012)]{bfak}Boyd, R.N., Famiano, M.A., Meyer, B.S., Motizuki, Y., 
Kajino, T., \& Roederer, I.U. 2012, \apj, 744, L14

\bibitem[Burbidge et al. (1957)]{b2fh}Burbidge, E.M., Burbidge, G.R., Fowler, W.A. \& Hoyle, R. 1957, Rev. Mod. Phys. 29, 547

\bibitem[Caballero, McLaughlin, \& Surman(2012)]{caballero12}Caballero, O.L., McLaughlin, G.C., \& Surman, R. 2012, \apj, 745, 170

\bibitem[Cescutti et al.(2006)]{travaglio}Cescutti, G., Francois, P., Matteucci, F., Cayrel., R., \& Spite, M.
2006, A\&A, 448, 557
%
\bibitem[Cristallo et al. (2015)]{cristallo15}Cristallo, S., Abia, C., 
Straniero, O., \& Piersanti 2015, \apj, 801, 53
%
\bibitem[Cyburt et al.(2010)]{cyburt10} Cyburt, R.H., Amthor, A.M., Ferguson, R., Meisel, Z., Smith, K., Warren, S.,
Heger, A., Hoffman, R.D., Rauscher, T., Sakharuk, A., Schatz, H., Thielemann, F.K., \& Wiesher, M. 2010, \apjs,
189, 240
%
\bibitem[Famiano et al. (2006)]{famiano06}Famiano, M. A., Liu, T., Lynch, W. G., Mocko, M., Rogers, A. M., Tsang, M. B., Wallace, M. S., Charity, R. J., Komarov, S., Sarantites, D. G., Sobotka, L. G., \& Verde, G. 2006, Phys. Rev. Lett.,
97, 052701
%
\bibitem[Famiano et al. (2013)]{famiano13}Famiano, M.A., Boyd,
R.N., Kajino, T., Meyer, B., Motizuki, Y., \& Roederer, I.
2013, J. Phys: Conf. Ser., 445, 012025
%
\bibitem[Farouqi et al.(2009)]{farouqi09}Farouqi, K., Kratz, K.-L., Mshonkina, L.I., Pfeiffer, B., Cowan, J.J., 
Thielemann, F.-K., \& Truran, J.W. 2009, \apj, 694, L49

\bibitem[Freiburghaus, Rosswog, \& Thielemann (1999)]{neutron}Freiburghaus, C., Rosswog, S., \& Thielemann, F.-K.
1999 \apj, 525, L121

\bibitem[Goreily, Bauswein, \& Janka (2011)]{goreily11}
Goriely, S., Bauswein, A., \& Janka, H.-T. 2011, \apj, 738, L32

\bibitem[Hidaka, Kajino, and Mathews (2015)]{hidaka15}Hidaka, J., Kajino, T., \& Mathews, G.J. 2015, \apj (submitted)

\bibitem[Horiuchi et al.(2014)]{horiuchi14}Horiuchi, S., Nakamura, K., Takiwaki, T., Kotake, K., \& Tanaka, M. 2014, MNRAS, 445, L99

\bibitem[Ishimaru et al.(2004)]{ishimaru}Ishimaru, Y., Wanajo, S., Aoki, W., \& Ryan, S.G. 2004 \apj, 600, L47

\bibitem[Ishimaru, Wanajo, \& Prantzos (2015)]{ishimaru15}
Ishimaru, Y., Wanajo, W., \& Prantzos, N. 2015, \apj, 804, L35
%
\bibitem[Izutani, Umeda, \& Tominaga (2009)]{izutani09}Izutani, N., Umeda, H., \& Tominaga, N.
2009, \apj, 692, 1517
%
\bibitem[Jacobson et al. (2015)]{jacobson15}
	Jacobson, H.R., Keller, S., Frebel, A., Casey, A.R., 
	Asplund, M., Bessell, M.S., Da Costa, G.S., Lind, K., 
	Marino, A.F., Norris, J.E., Pena, J.M., Schmidt, B.P.,
	Tisserand, P., Walsh, J.M., Yong, D., \& Yu, Q.
	2015, \apj, 807, 171
	
\bibitem[Lattimer (2012)]{lattimer12}
Lattimer, J.M. 2012, Ann. Rev. Nucl. Part. Sci., 62, 485

\bibitem[Lattimer \& Swesty(1991)]{lseos}Lattimer, J.M. \& Swesty, F.D. 1991, Nuc. Phys. A, 535, 331

\bibitem[Meyer \& Adams(2007)]{meyer07} Meyer, B.S. \& Adams, D.C. 2007, Meteor. and Plan. Sci. Suppl., 2007, 5215

\bibitem[Nakamura et al. (2015)]{nakamura15}Nakamura, K., et al.
2015, A\&A, 582, A34

\bibitem[O'Connor \& Ott (2010a)]{GRID_ref}O'Connor, E. \& Ott, C.D. 2010, Class. Quant. Grav. 27, 114103

\bibitem[O'Connor \& Ott (2011)]{oconnor11}O'Connor, E. \& Ott, C.D. 2011, \apj, 730, 70

\bibitem[Qian (2000)]{qian00}Qian, Y.-Z., 2000 \apj, 534, L67
%
\bibitem[Preston \& Sneden(2000)]{preston00}
Preston, G.W. \& Sneden, C. 2000, AJ, 120, 1014
%
%\bibitem[Tominaga et al. (2007)]{tominaga07}Tominaga, N., Maeda, K., 
%Umeda, H., Nomoto, K., Tanaka, M., Iwamoto, N., Suzuki, T., \& Mazzali, P.A. 2007 \apj, 657, L77

\bibitem[Qian \& Wasserburg(2008)]{qw}Qian, Y.-Z. \& Wasserburg, G.J. 2008 \apj, 687, 272
\bibitem[Shen et al.(2011)]{shenn}Shen, H., Toki, H, Oyamatsu, K., \& Sumiyoshi, K. 1998, Prog. Thoer. Phys., 100,
1013
\bibitem[Shibagaki et al. (2016)]{shibagaki15}Shibagaki, S., Kajino, T., Mathews, G.J., Chiba, S., Nishimura, S., \&
Lorusso, G. 2016, \apj, 816, 79
%
\bibitem[Spite et al.(2014)]{spite14}
Spite, M., Spite, F., Bonifacio, P., Caffau, E., Francois, P., \& Sbordone, L. 2014, A\&A, 571, A40
%
\bibitem[Steiner, Lattimer, \& Brown(2004)]{lattimer}Steiner, A.W., Lattimer, J.M., \& Brown, E.F. 2010, \apj,
722, 33
%
\bibitem[Suda et al.(2016)]{suda_prep}
Suda, T., et al. 2016 PASJ (submitted)
%
\bibitem[Suda et al.(2011)]{suda11}
 Suda T., Yamada S., Katsuta Y., Komiya Y., Ishizuka C., Aoki W., \&
 Fujimoto M.Y. 2011,
 Mon. Not. R. Astron. Soc., 412, 843
%
\bibitem[Suda et al.(2008)]{saga}Suda, T., Katsuta, Y., Yamada, S, Suwa, T., Ishizuka, C., Komiya, Y., Sorai, K., Aikawa, M.,
\& Fujimoto, M.Y. 2008, PASJ, 60, 1159

\bibitem[Sumiyoshi, Yamada, \& Suzuki (2008)]{sumiyoshi08}
Sumiyoshi, K., Yamada, S., \& Suzuki, H. 2008, \apj 688, 1176

\bibitem[Takahashi et al.(1994)]{takahashi94}Takahashi, K., Witti, J., \& Janka, H.-T. 1994, A\&A, 286, 857
%
\bibitem[Timmes, Woosley, \& Weaver(1995)]{timmes}Timmes, F.X., Woosley, S.E.,
\& Weaver, T.A. 1995, \apj, 98, 617
%
\bibitem[Travaglio et al. (2004)]{travaglio04}Travaglio, C., 
Gallino, R., Arnone, E., Cowan, J., Jordan, F., \& Sneden, C.
2004, \apj, 601 864

\bibitem[Wallerstein et al.(1997)]{wallerstein97}Wallerstein, G. et al. 1997, Rev. Mod. Phys., 69, 995
\bibitem[Wanajo et al. (2014)]{wanajo14}Wanajo, S., Sekiguchi, Y.,
Nishimura, N., Kiuchi, K., Kyutoko, K., \& Shibata, M. 
2014, \apj, 789, L39
\bibitem[Woosley et al.(1994)]{woo94}Woosley, S.E., Wilson, J.R., Mathews, G.J., Hoffman, R.D., \& Meyer, B.S. 1994, \apj, 433, 229

\bibitem[Warren et al. (2014)]{warren14}Warren, M.L., et al.
2014, Phys. Rev. D, 90, 103007

\bibitem[Woosley, Heger \& Weaver(2002)]{ww02}Woosley, S.E., Heger, A., \& Weaver, T.A. 2002, Rev. Mod. Phys., 74, 1015

\bibitem[Woosley \& Weaver (1995)]{ww95}
Woosley, S. E., \& Weaver, T. A. 1995, \apjs, 101, 181

\bibitem[Wu te al.(2014)]{wu14}Wu, M.-R., Fischer, T., Huther, L., et al. 2014, Phys. Rev. D, 89, 061303

\bibitem[Yamada et al. (2013)]{yamada13}
Yamada, S., Suda, T., Komiya, Y., Aoki, W., \& Fujimoto, M.Y.
2013, MNRAS, 436, 1362
\end{thebibliography}
\end{document}